\documentstyle[12pt,aaspp4]{article}
\newcommand    \be         {\begin{equation}}
\newcommand    \ee         {\end{equation}}
\newcommand    \Angstrom   {\,{\rm \AA}}          

\newcommand    \K          {\,{\rm K}}
\newcommand    \cm         {\,{\rm cm}}

\newcommand    \yr	   {\,{\rm yr}}
\newcommand    \nH         {n_{\rm H}}
\newcommand    \simlt      {\lesssim}
\newcommand    \simgt      {\gtrsim}
\newcommand    \gtsim      {\gtrsim}

\newcommand    \ta	   {\tau_{\rm{a}}}

\lefthead{Weingartner \& Draine}
\righthead{Depletion onto Very Small Grains}

\begin{document}

\title{Interstellar Depletion onto Very Small Dust Grains}

\author{Joseph C. Weingartner}
\affil{Physics Dept., Jadwin Hall, Princeton University,
        Princeton, NJ 08544, USA; josephw@astro.princeton.edu}

\and

\author{B.T. Draine}
\affil{Princeton University Observatory, Peyton Hall,
        Princeton, NJ 08544, USA; draine@astro.princeton.edu}

\begin{abstract}
	
We consider the depletion of elements from the interstellar gas onto a 
population of very small dust grains.  Adopting a grain model in which 
$\sim 4 \%$ of the cosmic C abundance is in grains with radii $\le 10 
\Angstrom$, we find that the rate of accretion onto these grains is
adequately fast to account for the observed large depletions of elements 
like Ti, without invoking unreasonably high rates of mass 
transfer between interstellar phases or low grain destruction rates.  
If these grains are composed of arene rings, then only a limited number of
metal atoms can be locked up in them.  The depletion would be quenched
when this limit is reached, unless there is a mechanism for transferring the
metals to larger grains and refreshing the
very small grain population, for example by grain coagulation and shattering
in the diffuse ISM.  If Fe depletes onto the very small grains, then for 
reasonable coagulation rates there is at least one metal atom per five C
atoms in the very small grain population.  Furthermore, $\approx 60 \%$ of 
the cosmic Fe is associated with
the carbonaceous grain population.  It is unclear 
whether this scenario is compatible with observations.  
However, if there is another population 
of very small grains, with a large capacity for holding Fe atoms, it 
might be the sink for the most heavily depleted elements.

\end{abstract}

\keywords{dust --- ISM: abundances}

\section{Introduction}

Ultraviolet and optical absorption line studies have shown that many elements
in the interstellar gas have lower abundances, relative to hydrogen, than in 
the Sun or in nearby B stars.  The depletion of species X is defined by 
\be
\delta(\rm{X}) \equiv \frac{{[n(\rm{X})/n_{\rm{H}}]}_{\rm{gas \; phase}}}
{{[n(\rm{X})/n_{\rm{H}}]}_{\rm{cosmic}}},
\ee
where $n(\rm{X})$ is the number density of species X and $n_{\rm{H}}$ is the
number density of hydrogen.
Savage \& Sembach (1996) have recently reviewed depletion measurements for
various interstellar environments.  

The conventional explanation for interstellar depletions is that some of the 
atoms have been accreted by dust grains.  However, efforts to quantify this
hypothesis for the most heavily depleted elements in diffuse gas
have encountered difficulties.  For 
example, Draine (1990) found that implausibly high rates of mass transfer 
from molecular clouds to the diffuse medium were required, but noted that the
inclusion of a population of very small grains might help.  The existence of
such a population is inferred from observations of $3$ to $60 \micron$
infrared emission, presumably generated by grains small enough to reach 
temperatures of 30 -- 300 K 
or more upon the absorption of a single starlight
photon (e.g. Draine \& Anderson 1985).  Very small grains are
expected to be more effective accretors than bigger ``classical'' grains for
two reasons.  First, they contribute more total surface area.  Second,
they tend to be negatively charged or neutral, whereas the
classical grains tend
to be positively charged.  Since the depleted metals in the gas are generally 
singly ionized, the collision cross sections for small grains are enhanced 
and those for big grains are diminished.

In this study, we repeat the analysis presented by Draine (1990), but
including a population of
very small grains.  In particular, we will aim to account for the depletion 
of Ti, which has been observed in both a cool diffuse cloud and a warm diffuse
cloud along the line of sight towards $\zeta$ Oph.  
For the cool cloud towards $\zeta$ Oph, $\delta({\rm Ti})
= 1 \times 10^{-3}$; for the warm cloud, $\delta({\rm Ti})= 5 \times 10^{-2}$  
(Savage \& Sembach 1996).  Although these particular clouds may not be typical,
the severe Ti depletions appear to be.  For sight lines along which the average
H number density is 3 cm$^{-3}$, $\delta({\rm Ti})
= 2 \times 10^{-3}$ (Jenkins 1989),
and Joseph \& Jenkins (1991) found that $\delta({\rm Ti})
< 4 \times 10^{-4}$ for the 
cool clouds towards $\pi$ Sco.  Although Ca is more severely depleted than Ti
towards $\zeta$ Oph, Crinklaw, Federman, \& Joseph (1994) 
conclude that generally Ca is
depleted to the same extent as Ti, as a function of average H number density
along the sight line.  
Depletion studies involving Ca are complicated by the fact that some Ca is
expected to be doubly ionized, but Ca$^{++}$ has no resonance lines at 
wavelengths longward of $912 \Angstrom$.

In \S 2 we derive expressions for the rate at which ions are removed from the 
gas by accreting onto grains, and compute accretion timescales for three 
interstellar phases.  In \S 3 we consider a scenario for mass exchange between
phases and derive the constraints which the observed Ti depletions impose
on the timescales for dust destruction and
mass transfer between phases.  We conclude in 
\S 4 by discussing the plausibility of the resulting scenario.

\section{Depletion Rates}

\subsection{Rate Coefficients}

Ions arrive at the surface of a spherical grain at a rate
\be
R_{\rm{arr}} = \int_0^{\infty} dv {\left(\frac{2}{\pi}\right)}^{1/2}
{\left(\frac{m}{kT}\right)}^{3/2} v^2 \exp \left( - \frac{m v^2}{2 k T}
\right)
 \pi b_{\rm{max}}^2 (v) v n,
\label{arrive}
\ee
where $k$ is Boltzmann's constant, $T$ is the gas temperature, $m$ is the
ion mass, $b_{\rm{max}}$ is the maximum
impact parameter for which an ion strikes the grain, and $n$ is the ion 
number density.  When the grain is charged the ion moves in a Coulomb 
potential for which (Spitzer 1941)
\be
b_{\rm{max}}(v) = \cases{0 &for ${mv^2\over 2} < {Z_g Z_i e^2\over a}$\cr
a\left(1-{2Z_g Z_i e^2\over mav^2 }\right)^{1/2}
&for ${mv^2\over 2} > {Z_g Z_i e^2\over a}$\cr} \  , 
\ee
where $Z_g e$ is the grain charge, $Z_i e$ is the ion charge, $m$ is the ion 
mass, and $a$ is the grain radius.  For charged grains,
we neglect the modifications to 
$b_{\rm{max}}$ which result from the ``image charge'' contribution to the 
interaction (Draine \& Sutin 1987).  Neutral grains are polarized
by the ion's electric field.  Treating the grain as a perfect conductor,
Draine \& Sutin (1987) found that
\be
b_{\rm{max}}(v) 
= a\left[1 + \left({4Z_i^2e^2\over m v^2 a}\right)^{1/2}\right]^{1/2}~~~.
\ee

The accretion timescale, $\ta$, is defined and computed from
\be
\ta^{-1} \equiv -\frac{1}{n} \frac{dn}{dt} = {\rm{A}}^{-1/2} s {\left( \frac
{8 \pi k T}{m_{\rm{p}}}\right)}^{1/2} \int  da \, a^2 
\frac{dn_{\rm{gr}}}{da} D(a),
\label{tacc}
\ee
where $\rm{A}$ 
is the mass number of the ion, $s$ is the sticking coefficient of
the ion on the grain, $m_{\rm{p}}$ is the proton mass,
and $n_{\rm{gr}}(a)$ is the number 
density of grains with radii less than $a$.  
The enhancement 
factor $D(a)$ depends on $Z_i$ and $Z_g$.  With $Z_i$ fixed,
\be
D(a) = \sum_{Z_g} f(Z_g, a) B(Z_g, a),
\ee
where (Draine \& Sutin 1987)
\be
B(Z_g, a) = \cases{\exp\left(\frac{-Z_g Z_i e^2}{k T a}\right) &for $Z_g Z_i 
> 0$\cr
\left(1-\frac{Z_g Z_i e^2}{k T a}\right) &for $Z_g Z_i < 0$\cr
1+{\left(\frac{\pi Z_i^2 e^2}{2 k T a}\right)}^{1/2} &for $Z_g = 0$\cr} \ .
\ee
The grain charge distribution function, $f(Z_g, a)$,
is obtained using the method described
by Weingartner \& Draine (1998).
In Figure 1 we plot $D(a)$, with $Z_i = 1$, for graphite and
silicate grains in three interstellar environments:
cold neutral medium (CNM),  warm neutral medium (WNM), and warm ionized 
medium (WIM).  The adopted
hydrogen number density $\nH$, gas temperature $T$, and
ratio of electron density to hydrogen density $x_e$, for these three 
environments, and for molecular clouds (MC), are displayed in Table \ref{tph1}.
Since photoelectric emission is 
important in the grain charging, a radiation field must be specified; we 
used the spectrum of Mathis, Mezger, \& Panagia (1983). 
The absorption cross sections were calculated using
a Mie theory code derived from BHMIE (Bohren \& Huffman 1983), 
with dielectric functions as described by Draine \& Lee (1984) and
Laor \& Draine (1993).  In molecular clouds, the grains are predominantly 
neutral and $Z_i = 0$; thus $D(a) \approx 1$.

\subsection{Grain Population}

We consider two model grain size distributions, both of which are extensions
of the MRN distribution (Mathis, Rumpl, \& Nordsieck 1977),
\be
dn_{gr} = C a^{-3.5} \nH da, \quad a_{\rm{min}} < a < a_{\rm{max}},
\ee
with coefficients given by Draine \& Lee (1984): $C = {10}^{-25.13} {\cm}^
{3.5}$ for graphite and $C = {10}^{-25.11} {\cm}^{3.5}$ for silicate.  The 
MRN distribution runs from $a_{\rm{min}} = 50 \Angstrom$ to 
$a_{\rm{max}} = 0.3 \micron$, but for the 
graphite component we extend the lower limit to $4 \Angstrom$, about the 
minimum size expected to be stable against sublimation in the 
interstellar radiation
field (Guhathakurta \& Draine 1989).  We do not similarly extend the 
silicate component, because the observed
IR emission from diffuse clouds does not exhibit the $10 \micron$ silicate 
feature (Mattila et al. 1996; Onaka et al. 1996).  
For our first distribution ({\it i}), we retain the MRN power law
throughout the extended size range.
If graphite grains have a density of $2 \, \rm{g} {\cm}^{-3}$, then 
the grains with $a \le 10 \Angstrom$ ($\le 420$ C atoms), contain $\sim 2 \%$
of the cosmic C abundance.  This is conservative, as attempts to model the
observed IR emission (e.g. D\'{e}sert, Boulanger, \& Puget 1990)
have suggested values two or three times bigger.  Therefore, for 
our second distribution ({\it ii}), we adopt for the graphite component
the form suggested by Draine \&
Anderson (1985), which differs from ({\it i}) only for $a < 30 \Angstrom$,
where
\be
dn_{gr} = C {(30 \Angstrom)}^{0.5} a^{-4.0} \nH da.
\ee 
In this distribution, grains with $a \le 10 \Angstrom$ contain $\sim 4\%$
of the cosmic C abundance.  The quantity
\be
P(a) \equiv {\left(\frac{\rm{A}}{48}\right)}^{1/2}
\frac{1}{s}\frac{d\tau_a^{-1}}{d\ln a}
\ee
indicates how grains of different size contribute to the 
overall accretion rate.
In Figures 2 and 3 we show $P(a)$ for $Z_i = +1$ ions and CNM,
WNM, and WIM conditions.  Clearly, 
the very small grains completely dominate the accretion.

\section{Consequences}

\subsection{Model for Mass Exchange}

We consider mass exchange between the following three gaseous environments:  
warm, cold, and molecular.  Hot ionized gas is neglected because of its small
mass fraction.  We also consider the removal of mass from the ISM 
during the star formation process, and the addition of mass to the ISM from 
evolved stars (stellar winds and supernovae)
and from the infall of extragalactic material.  For 
simplicity, we assume that the total mass and the metallicity of the ISM 
remain constant in time.  To properly account for changes in metallicity
with time, it would be necessary to consider the dependence of 
the dust abundance
on metallicity and the distribution of newly formed metals over the 
interstellar phases.  
The fraction of mass removal due to star formation
occurring in phase i is denoted $p_{\rm{i}}$, where i $= $ c, w, or m for 
cold, warm, or molecular, respectively; 
we assume $p_{\rm{m}} = 1$ and $p_{\rm{c}}=p_{\rm{w}}=0$.
We consider mass addition (both from evolved stars and infall)
only into the warm and cold media, with
the same relative contributions from the two sources in both media.  Then the
mass input can be treated as if it were derived from a single source, with Ti
depletion ${\delta}_{\rm{in}}$; if there is no Ti in the accreted extragalactic
material, then ${\delta}_{\rm{in}}$ has the same value as the depletion in the
stellar outflow (relative to its enhanced metallicity).  The timescale for 
injection of the entire mass of the ISM is denoted by ${\tau}_{\rm{in}}$, and
the fraction of the input that goes into phase i is denoted by
$q_{\rm{i}}$.

Ti is removed from the gas by accretion
onto very small grains, with timescales $\tau_{\rm{a,i}}$.
Since we wish to explain the
most heavily depleted elements, we make the extreme assumption that atoms are
only returned to the gas from grains when
the grains are destroyed.  We denote the timescale for the destruction of all
the grains in phase i as $\tau_{\rm{d,i}}$ and the mass fraction in phase i  
as $f_{\rm{i}}$.  In the warm and cold phases, 
grains are destroyed principally in supernova shock waves.  The timescale for
destroying all the grains in the ISM in these shocks is denoted $\tau_
{\rm{d}}$; the fraction of this destruction occurring in the cold medium is 
denoted $g_{\rm{c}}$ [$f_{\rm{c}} \tau_{\rm{d,c}}^{-1} = g_{\rm{c}} 
\tau_{\rm{d}}^{-1}$ and $f_{\rm{w}} \tau_{\rm{d,w}}^{-1} = (1-g_{\rm{c}}) 
\tau_{\rm{d}}^{-1}$].
We will suppose a small amount of additional grain destruction to take place
in molecular clouds, with a timescale $\tau_{\rm{d,m}}$.

\subsection{Steady State Equations}

We suppose the following mass transfers occur:  from warm to 
cold, cold to warm, cold to molecular, and molecular to warm, as indicated in  
Figure 4.  We define the
transfer timescales $\tau_{\rm{ij}}$ as the time to transfer the entire mass 
of phase i to phase j.  Assuming the depletions in each phase to remain 
constant yields
\be
\frac{1-\delta_{\rm{j}}}{\tau_{\rm{d,j}}} + \sum_{k \not= j} \frac{f_{\rm{k}}}
{f_{\rm{j}}} \frac{1}{\tau_{\rm{kj}}} \left(\delta_{\rm{k}} - 
\delta_{\rm{j}}\right) + \frac{q_{\rm{j}}}{f_{\rm{j}}} \frac{1}
{\tau_{\rm{in}}} \left( {\delta}_{\rm{in}} - \delta_{\rm{j}} \right)
= \frac{\delta_{\rm{j}}}{\tau_{\rm{a,j}}},
\ee
where
$\delta_{\rm{i}}$ denotes the Ti depletion in phase i (McKee 1989; Draine 
1990).  Constant mass in each phase yields
\be
\sum_{k \not= j} \left( \frac{f_{\rm{j}}}{\tau_{\rm{jk}}} - \frac
{f_{\rm{k}}}{\tau_{\rm{kj}}} \right) = \frac{1}{\tau_{\rm{in}}} \left(
q_{\rm{j}} - p_{\rm{j}} \right).
\ee

We treat 
$\delta_{\rm{c}}$, $\delta_{\rm{w}}$, $\tau_{\rm{a,c}}$,
$\tau_{\rm{a,w}}$, $\tau_{\rm{a,m}}$,
$\tau_{\rm{in}}$, $\delta_{\rm{in}}$, $q_{\rm{c}}$, 
$q_{\rm{w}} = 1 - q_{\rm{c}}$, $g_{\rm{c}}$, $g_{\rm{w}} = 1 - g_{\rm{c}}$,
$\tau_{\rm{d,m}}$, and the mass fractions $f_{\rm{j}}$ as known quantities
determined by observation ($f_{\rm{j}}$,
$\delta_{\rm{c}}$, $\delta_{\rm{w}}$), the adopted grain model 
($\tau_{\rm{a,c}}$, $\tau_{\rm{a,w}}$, $\tau_{\rm{a,m}}$), the model for mass
injection ($\tau_{\rm{in}}$, $\delta_{\rm{in}}$, $q_{\rm{c}}$,
$q_{\rm{w}}$), the relative rates for grain destruction in the different
phases ($g_{\rm{c}}$, $g_{\rm{w}}$), and the assumed rate for grain 
destruction in molecular clouds ($\tau_{\rm{d,m}}$).  
Recall that we assume all star formation to occur in molecular clouds
($p_{\rm{m}} = 1$, $p_{\rm{c}} = p_{\rm{w}} = 0$). 
There are then five equations and six unknown quantities:  the grain
destruction timescale $\tau_{\rm{d}}$, the depletion in molecular clouds
$\delta_{\rm{m}}$, and the four transfer timescales ($\tau_{\rm{wc}}$,
$\tau_{\rm{cm}}$, $\tau_{\rm{cw}}$, $\tau_{\rm{mw}}$).
If $\tau_{\rm{d}}$ is also specified, the solution
for the remaining variables is as follows:
\be
\delta_{\rm{m}} = \left\{\frac{1}{\tau_{\rm{d}}}\left[\left(
1-\delta_{\rm{w}}\right)
+ g_{\rm{c}}\left(\delta_{\rm{w}}-\delta_{\rm{c}}\right)\right]
 + \frac{f_{\rm{m}}}
{\tau_{\rm{d,m}}}-\frac{f_{\rm{w}} \delta_
{\rm{w}}}{\tau_{\rm{a,w}}} -\frac{f_{\rm{c}} \delta_{\rm{c}}}{\tau_{\rm{a,c}}}
+ \frac{\delta_{\rm{in}}}{\tau_{\rm{in}}}\right\}{\left(\frac{f_{\rm{m}}}{\tau_
{\rm{a,m}}}+ \frac{f_{\rm{m}}}{\tau_{\rm{d,m}}} + \frac{1}{\tau_{\rm{in}}}
\right)}^{-1},
\label{dm}
\ee
\be
\tau_{\rm{wc}} = f_{\rm{w}} \left( \delta_{\rm{w}}-
\delta_{\rm{c}} \right) {\left[ \frac{f_{\rm{c}}\delta_{\rm{c}}}
{\tau_{\rm{a,c}}}-\frac{g_{\rm{c}} \left(1-\delta_{\rm{c}}\right)}{\tau_
{\rm{d}}}
-\frac{q_{\rm{c}} \left(\delta_{\rm{in}} - \delta_{\rm{c}} \right)}
{\tau_{\rm{in}}}\right]}^{-1},
\ee
\be
\tau_{\rm{cm}} = \frac{f_{\rm{c}}}{f_{\rm{m}}} 
\left(\delta_{\rm{c}} - \delta_{\rm{m}}\right)
{\left( \frac{\delta_{\rm{m}}}{\tau_{\rm{a,m}}} -\frac{1-\delta_{\rm{m}}}
{\tau_{\rm{d,m}}}\right)}^{-1},
\ee
\be
\tau_{\rm{mw}} = {\left[\frac{1}{\delta_{\rm{c}}-\delta_{\rm{m}}}
\left( \frac{\delta_{\rm{m}}}{\tau_{\rm{a,m}}} -\frac{
1-\delta_{\rm{m}}}{\tau_{\rm{d,m}}}\right)
-\frac{1}{f_{\rm{m}} \tau_{\rm{in}}}\right]}^{-1},
\label{taumw}
\ee
and
\be
\tau_{\rm{cw}} = f_{\rm{c}} \left(\delta_{\rm{w}}-\delta_{\rm{c}}\right)
{\left[\frac{f_{\rm{c}} \delta_{\rm{c}}}{\tau_{\rm{a,c}}} -\frac{g_{\rm{c}}
\left(1-\delta_{\rm{c}}\right)}{\tau_{\rm{d}}} - f_{\rm{m}}
\frac{\delta_{\rm{w}} -
\delta_{\rm{c}}}{\delta_{\rm{c}}-\delta_{\rm{m}}} \left( \frac{
\delta_{\rm{m}}}{\tau_{\rm{a,m}}} - \frac{1-\delta_{\rm{m}}}
{\tau_{\rm{d,m}}}\right) + \frac{q_{\rm{c}} \left( \delta_{\rm{w}} -
\delta_{\rm{in}}\right)}{\tau_{\rm{in}}}
\right]}^{-1}.
\label{taucw}
\ee

The destruction timescale $\tau_{\rm{d}}$ is constrained by requiring the 
above quantities to be positive.  For example,
we assume $\delta_{\rm{w}} \ge \delta_{\rm{c}}$; then by demanding  
$\tau_{\rm{wc}} \ge 0$ we find 
\be
\tau_{\rm{d}} \ge g_{\rm{c}} \left(1-\delta_{\rm{c}}\right) {\left[ \frac
{f_{\rm{c}} \delta_{\rm{c}}}{\tau_{\rm{a,c}}} - \frac{q_{\rm{c}} \left(
\delta_{\rm{in}} - \delta_{\rm{c}}\right)}{\tau_{\rm{in}}} \right]}^{-1}.
\label{tde1}
\ee
Requiring $\delta_{\rm{m}} \ge 0$ implies
\be
\tau_{\rm{d}} \le \left[\left(1-\delta_{\rm{w}}\right) + g_{\rm{c}} \left(
\delta_{\rm{w}} - \delta_{\rm{c}} \right) \right]
{\left( \frac{f_{\rm{w}} \delta_{\rm{w}}}
{\tau_{\rm{a,w}}} + \frac{f_{\rm{c}} \delta_{\rm{c}}}{\tau_{\rm{a,c}}}
-\frac{f_{\rm{m}}}{\tau_{\rm{d,m}}}
- \frac{\delta_{\rm{in}}}{\tau_{\rm{in}}}\right)}^{-1}.
\label{tde2}
\ee

\subsection{Exemplary Solutions}

The mass of the ISM is estimated to be $\sim 5 \times 10^9 \, 
{\rm{M}}_{\odot}$,
and the star formation rate (or rate of infall plus stellar outflow) is 
$\sim 5 \, {\rm{M}}_{\odot} \, {\rm{yr}}^{-1}$ (Timmes, Diehl, \& Hartmann
1997 and references therein); 
hence we adopt $\tau_{\rm{in}} \approx 10^9 \,\rm{yr}$.
In Table \ref{tph2} we show the adopted values of $\tau_{\rm{a,i}}$ (for each
grain size distribution),
$f_{\rm{i}}$, and $\delta_{\rm{i}}$; the accretion timescales were computed 
using (\ref{tacc}), with $\rm{A}=48$ and $s=1$.  For the 
warm and cold media we take $Z_i = +1$, while for molecular clouds we take
$Z_i = 0$, as well as raising $a_{\rm{min}}$ to $150 \Angstrom$.  
The adopted values for the mass fractions $f_{\rm{i}}$ are rather uncertain,
and not necessarily constant in time.  

In Table \ref{solns} we present three exemplary solutions.  
Case A uses distribution ({\it i}), with $\delta_{\rm in}=0.1$,
and cases B and C assume distribution ({\it ii}), with
$\delta_{\rm in}=0.1$ (case B) and 1 (case C).
For each case the adopted value of $\tau_{\rm d}$ lies at the center
of the allowed range of values, which
generally has a width $ \simlt 10^7 \, \rm{yr}$.  

The results are quite insensitive to $q_{\rm{c}}$ (the fraction of infall  
entering the cool phase), which we set to $q_{\rm{c}} = 0.1$, and
to $g_{\rm{c}}$ (the fraction of grain destruction taking place in the CNM),
which we set to $g_{\rm{c}} = 0.05$.  Only $\delta_{\rm{m}}$ is sensitive
to $\tau_{\rm{d,m}}$, which we set to $1 \times 10^{10} \, \rm{yr}$, and to   
$\tau_{\rm{a,m}}$ (although $\tau_{\rm{cm}}$ and $\tau_{\rm{mw}}$ begin to 
increase significantly when $\tau_{\rm{a,m}}$ gets so large that $\delta_
{\rm{m}} > \delta_{\rm{c}}$, but this regime is unrealistic).
For distribution ({\it i}), 
$\tau_{\rm{d}} = 1.4 \times 10^9 \, \rm{yr}$ when $\delta_{\rm{in}} = 0$ and 
increases with increasing $\delta_{\rm{in}}$; above $\delta_{\rm{in}} = 0.72$,
there is no solution (although if $\tau_{\rm{in}}$ is increased to $1.4
\times 10^{10} \, \rm{yr}$, there are solutions up through $\delta_{\rm{in}}
 = 1.0$).  
For ({\it ii}), $\tau_{\rm{d}}$ increases from $6.3
\times 10^8 \, \rm{yr}$ for $\delta_{\rm{in}} = 0$ to $1.6 \times 10^9 
\, \rm{yr}$ for $\delta_{\rm{in}} = 1$.

\section{Discussion}

\subsection{Overall Timescales and Depletions}

Jones and collaborators have made a series of models for grain destruction in
shocks (Jones et al. 1994, 1996).  They find that the 
smallest grains are the most resilient, since most destruction results from 
nonthermal sputtering off betatron-accelerated grains, and the larger grains
are more effectively accelerated by the betatron mechanism.
Less frequent, faster shocks efficiently destroy
smaller grains through thermal sputtering.  Furthermore, they find that 
shattering in collisions between larger grains produces large numbers of 
very small grains.  Presumably these very small 
grains later coagulate to form bigger 
grains, maintaining the overall grain distribution.  
Since atoms which have accreted onto very small grains are incorporated 
into larger grains via coagulation, it would seem 
most appropriate for our purpose to adopt the timescale for destruction of 
the entire distribution.  Jones et al. (1996) estimate the 
lifetime of carbonaceous
grain material throughout all the phases of the interstellar 
medium, $\tau_{\rm{d}} = 6 \times 10^8 \, \rm{yr}$, which 
about equals the value adopted in \S 3 for case B, but is 
about a factor of 2 smaller than the values adopted for our other exemplary 
cases.  The derived transfer timescales are in the 
expected range (McKee 1989).  Thus, our case B example shows that
accretion onto very small 
carbonaceous grains can occur fast enough to result in the depletions of 
even the most heavily depleted elements, if they stick perfectly.

\subsection{Sticking Coefficients}

The ``sticking'' coefficient $s = p_1 p_2$, where $p_1$ is the probability 
that the impinging ion (e.g. Ti$^+$) will be trapped on the grain surface
long enough ($\simgt 10^{-12} \, \rm{s}$) to get rid of its kinetic energy and 
``thermalize'', and $p_2$ is the probability that the atom will then
remain on the
grain for $\simgt 10^8 \, \rm{yr}$, until the grain is overrun by a 
supernova blastwave.  This requires that the atom not be removed by either
photodesorption (by a UV photon) or chemical reaction (with impinging H or O,
in particular).  
For cold ($T \simlt 20 \, \rm{K}$) grains in cold gas ($T \simlt 100 \, 
\rm{K}$), 
it seems likely that $p_1$ is of
order unity for species other than the inert gases.  

The probability $p_2$ is
most likely highly variable from one species to another.  We would require 
$p_2$ to be of order unity for Ti and other species that are strongly   
depleted, and $p_2 \ll 1$ for species like N and S, which are not depleted.  
Photodesorption can keep a grain surface ``clean'' of species which are
susceptible to photodesorption (Draine \& Salpeter 1979), since the
lifetime against photodesorption by ambient starlight could be as short
as $\sim3000\yr$ for an atom or molecule.
Efficient photodesorption has been observed for UV-irradiated H$_2$O ice
(Westley et al. 1995, 1996), but the 
photodesorption cross sections for various metals at the
various bonding sites on very small carbonaceous grains are not known.
The small (carbonaceous) grains might consist of  
layers of fairly extended arene sheets. In this case, an atom located between 
sheets may have its bond(s) broken by a photon, but then transfer its excess
energy to the grain through collisions, and promptly form new bonds.  Thus,
variations in $p_2$ from element to element might arise from different 
cross sections for bond breaking or from different abilities to diffuse out of
the grain once bonds are broken. 

\subsection{Saturation of Very Small Grains with Metals}

Another consideration bearing on the sticking of metals regards the 
finite number of bonding sites in the very small carbonaceous grain
population.  Klotz et al. (1995) noted that in the most common PAH complexes,
each metal is bonded by six $p_z$ electrons on the same ring, implying that 
rings adjacent to a metal-bonded ring are not available to bond other metal 
atoms.  They concluded that there can be at most one metal per two rings, or
12 C atoms.  However, there are many different possible bonding modes
(Maslowsky 1993), so we shall be more flexible in our criterion for what
constitutes a ``full'' grain.  Suppose fast accretion can be maintained so
long as there are at most $\beta$ metal atoms per C in the very small 
carbonaceous grains.  (Klotz et al. would only consider $\beta
\le 1/12$.)  Then, the very small grains become saturated when their metal
content reaches $\beta$ and must be 
replaced by a fresh population of very small carbonaceous grains which
are largely metal-free, and therefore capable of accreting metals.
We will refer to this as ``refreshment'' of the
small grain population, and will let $\tau_{\rm re}$ denote the timescale
for this refreshment of the small carbonaceous grains.
This refreshment would presumably occur via coagulation of very small grains
onto classical grains, balanced by the shattering of classical grains to
create ``fresh'' very small grains.  
If the coagulation takes place in
molecular clouds and the shattering in the warm medium, then the refreshment
rate $\tau_{\rm{re}}^{-1} < \tau_{\rm{cm}}^{-1}$, the rate for 
exchange of mass from CNM to molecular clouds.
In this scenario, rapid refreshment would require rapid mass exchange
between the molecular clouds and other phases.
Alternatively, both shattering and coagulation might occur in the CNM.
In this case, we demand that $\tau_{\rm{re}}
\ge \tau_{\rm{coag}}$, where the timescale for grains to exit the very small
grain population by coagulation is roughly approximated by
\be
\tau_{\rm{coag}} \approx {\left(\pi v \int a^2 \frac{dn_{\rm{gr}}}{da} da
\right)}^{-1}
\approx
1 \times 10^{7} \, {\rm{yr}} \left( \frac{{\rm{km}} \, {\rm{s}}^{-1}}{v}
\right) \left(\frac{30 {\rm{cm}}^{-3}}{n_{\rm{H}}}\right).
\ee
The evaluation is for grain distribution ({\it ii}).  Interstellar 
turbulence and radiation pressure could conceivably
lead to relative speeds 
between grains $v \sim 1 \, \rm{km} \, {\rm{s}}^{-1}$,
although estimated drift speeds resulting from anisotropic radiation
tend to be about a factor of 50 smaller (Weingartner \& Draine 1998).

If there are $\gamma$ metal atoms per C atom in the entire carbonaceous 
grain population, then freshly shattered very small grains  will
presumably have $ \sim \gamma$
metals per C as well.  Thus, grains of size $a$ remain unsaturated if
\be
\tau_{\rm{re}}(a) \le \tau_{\rm{re}}^{\rm{max}}(a) = (\beta - \gamma)
\frac{N_{\rm{C}}(a)}{R_{\rm{st}}(a)},
\ee
where $N_{\rm{C}}(a)$ is the number of C atoms in a grain and $R_{\rm{st}}(a)$
is the rate at which metals stick to the grain.
We express $\tau_{\rm{re}}^{\rm{max}}(a)$ 
in terms of its value for $a = 10 \Angstrom$:
\be
\frac{1}{\tau_{\rm{re}}^{\rm{max}}(a)} = 
\frac{D(a)}{D(10 \Angstrom)} \left( \frac{10
\Angstrom}{a} \right) \sum_{\rm{M}} \frac{1}{{\tau_{\rm{re}}^{\rm{max}}(10
\Angstrom)}_{\rm{M}}} \ ,
\ee
where the sum is over all metals M which accrete onto the very small grains.
For the CNM,
\be
{\tau_{\rm{re}}^{\rm{max}}(10\Angstrom)}_{\rm{M}} = 
11.6 \, {\rm{yr}} \, (\beta -
\gamma) {\left[\delta_{\rm{c}} x_{\rm{cos}} {\left(\frac{48}{A}\right)}^{1/2}
s \right]}^{-1}_{\rm{M}},
\label{tre}
\ee
where $x_{\rm{cos}}$ is an element's cosmic abundance, relative to H.
To approximate $\gamma(\rm{M})$, the contribution to $\gamma$ from metal M,
we assume that M is not at all 
incorporated into the carbonaceous grains during their formation processes.
Balancing the 
addition of M (by accretion in the CNM, with timescale $\tau_{\rm{a,c}}$) 
against the removal of M (during grain destruction, with timescale $\tau_ 
{\rm{d}}$, and star formation, with timescale $\tau_{\rm{in}}$), we find
\be
\gamma({\rm{M}}) = {\left(
\frac{f_{\rm{c}} \delta_{\rm{c}} x_{\rm{cos}}}{\tau_
{\rm{a,c}}} \right)}_{\rm{M}} \, \frac{m_{\rm{C}}}{m_{\rm{gra}}} {\left(
\frac{1}{\tau_{\rm{in}}} + \frac{1}{\tau_{\rm{d}}} \right)}^{-1},
\label{gm}
\ee
where $m_{\rm{gra}}$ is the mass of the entire carbonaceous grain population,
per H atom.

In Table \ref{compt} we display 
${\tau_{\rm{re}}^{\rm{max}}(10 \Angstrom)}_{\rm{M}}/(\beta
- \gamma)$, and $\gamma(\rm{M})$ for several elements, adopting
$m_{\rm{gra}} = 5.3 \times 10^{-27} \rm{g} / \rm{H}$. 
The values of $\tau_{\rm{a,c}}$ to be used in equation (\ref{gm}) 
are determined from equations (\ref{tde1}) and (\ref{tde2})
by demanding that $\tau_{\rm{d}} \approx 6 \times 10^8 
\, \rm{yr}$, and are denoted $\tau_{\rm{a,c}}^{\rm{req}}$ in Table 4;  
the values of $(48/A)^{1/2} s$ for use in equation 
(\ref{tre}) follow
from these required values of $\tau_{\rm{a,c}}$, assuming distribution   
({\it ii}).  Some metals might accrete
onto only the silicates, resulting in the growth of these grains, and not
contribute to the need to refresh the very small carbonaceous grains.  
With $(48/A)^{1/2} s = 1$, and adopting the graphite form of 
distribution ({\it ii}) for the
silicates, except with lower cutoff at $15 \Angstrom$ ($10 \Angstrom$),
$\tau_{\rm{a,c}} = 2 \times 10^6 \, \rm{yr}$ ($8 \times 10^5 \, \rm{yr}$).
Thus Mg and Si could possibly only accrete onto the silicates. 
Iron is a marginal case; however, it would be strange if Fe didn't
stick well to the very small grains when Ti and Ni do.  
Excluding Mg and Si, but including Fe, we find $\gamma = 8.3 \times 10^{-2}$
and $\tau_{\rm{re}}^{\rm{max}}(10\Angstrom)/(\beta - \gamma) = 
2.9 \times 10^8 \, 
\rm{yr}$.  Setting $\tau_{\rm{re}}(a) \ge \tau_{\rm{coag}} \gtsim
1 \times 10^7 \, \rm{yr}$, 
we find the minimum possible value for $\beta$ for which 
grains with size $a$ can be effectively refreshed.  This ranges from 
$\approx 0.35$ for $a = 4 \Angstrom$ to $\approx 0.12$ for $a = 10 \Angstrom$.

Thus, if Fe depletion is due to accretion onto very small carbonaceous grains,
then Fe is a major component in the composition of these grains, and $\approx
60 \%$ of the cosmic Fe resides in the entire carbonaceous grain population.
Although there appears to be no reason to exclude this possibility, it is 
certainly unexpected.  The alternative, that Fe depletes by accreting onto 
silicate grains, is unattractive: (1) it isn't clear that the accretion rate
onto the silicates is fast enough, and (2) one would expect Fe to stick to 
arene grains if Ti does.

\subsection{Another grain population?}

Perhaps there is a population of non-carbonaceous
very small grains, onto which 
the most heavily depleted elements accrete.
Tielens (1998) has recently inferred that the timescale for the return of 
Si from grains to gas is about an order of magnitude less than that for Fe,
since Si is much less depleted in the warm phase.  In computing our Table 4,
we took the same $\tau_{\rm{d}}$ for each species, resulting in different 
transfer timescales, as exemplified by $\tau_{\rm{mw}}$, the time for 
transfering gas from molecular clouds to the warm phase.  Of course,  
$\tau_{\rm{mw}}$ should be the same for each species.  If we take $\tau_
{\rm{a,c}} = 3 \times 10^{6} \, \rm{yr}$ for Si, appropriate for accretion
onto an MRN silicate distribution extended down to $15 \Angstrom$, then we
find $\tau_{\rm{d}} = 1.4 \times 10^8 \, \rm{yr}$ and $\tau_{\rm{mw}} = 
7 \times 10^7 \, \rm{yr}$.  The same $\tau_{\rm{mw}}$ is obtained for Ti and
Fe if their accretion timescales are increased by slightly less than a factor
of two, and $\tau_{\rm{d}} = 1.0 \times 10^9 \, \rm{yr}$.  Thus, we can 
imagine a silicate grain population deficient in Fe 
and with a destruction
timescale somewhat less than the $4 \times 10^8 \, \rm{yr}$ derived by 
Jones et al. (1996), along with a more resilient, Fe-rich grain population 
which includes very small grains.  
Metallic Fe grains are ruled out because they would produce excessive
magnetic dipole emission at 90 GHz (Draine \& Lazarian 1999).
Sembach \& Savage (1996) have suggested
oxides as a grain component, on the basis of gas phase abundance measurements
in diffuse halo clouds.  They found more Fe plus Mg per Si in dust
than expected for pure olivines or pure pyroxenes.  

A 20$\micron$ emission feature has recently
been detected in emission in warm circumstellar dust
shells, and attributed to FeO grains (Waters 1998).
Since single-photon heating of the very small interstellar
grains results in emission at
$\lambda \simgt 5 \micron$, candidate materials will be constrained by the 
observed emission spectrum of diffuse clouds.  Koike et al. (1981) and 
Henning et al. (1995) have experimentally determined IR optical constants
for well-known iron oxides, and reported $Q_{\rm{abs}}/a$ for spheres in the
Rayleigh limit ($Q_{\rm{abs}}$ is the absorption efficiency factor).  We 
reproduce their band positions in Table \ref{irobs}.  It is important to note
that the band positions depend on the particle shape.  Henning et al. (1995)
also calculated absorption for two continuous distributions of ellipsoids and 
found that the FeO
band position shifts from $19.9 \micron$ to $21.0 \micron$ or
$23.4 \micron$.  Serna, Oca\~{n}a, \& Iglesias (1987) 
and Wang, Muramatsu, \& Sugimoto (1998) discussed the 
effect of particle shape on the hematite spectrum.  We suspect that the 
$9 \micron$ feature in the Koike et al. hematite spectrum is due to an 
impurity in their sample; this feature is absent in the Nyquist \& 
Kagel (1971) spectrum and present, but attributed to an impurity, in the 
Sadtler spectrum (Ferraro 1982).  Thus, the Matilla et al. (1996) and 
Onaka et al. (1996) studies do not rule out very small grains with iron oxide
composition, since the longest wavelength they observed is $11.7 \micron$.
Spectrophotometry of the emission from diffuse clouds
in the 15 to $30 \micron$ range 
would resolve the issue, but are beyond current capabilities.
Thus at this time we cannot exclude the possibility that a significant
fraction of interstellar Fe resides in ultrasmall Fe oxide particles.

\subsection{Conclusion}

Observations of infrared emission in the 3 to $60 \micron$ range 
indicate that 
there is a population of very small dust grains, extending down to sizes of
a few $\Angstrom$.  The presence of several spectral features associated with
polycyclic aromatic hydrocarbons and the absence of the $10 \micron$
silicate feature imply that there are carbonaceous grains with 
$a < 10 \Angstrom$, but no such silicate grains.  
The rate at which gas cations collide with these ultrasmall grains
greatly exceeds that for collisions with bigger ``classical'' 
($a \gtsim 0.01\micron$) grains.
If the elements which are most heavily depleted in the ISM stick
efficiently upon collision, then accretion onto very small grains occurs
quickly enough to account for their depletions.  This fast accretion
is completely independent of the model adopted for the classical grains,
which is currently controversial (Mathis 1996).
Efficient sticking requires resistance against photodesorption in the 
interstellar radiation field and
against chemical attack.  Furthermore, if the depletion of the more
abundant element, Fe, is due to accretion onto very small carbonaceous grains,
then Fe would be a major component in the composition of these grains,
at a level 
which might be chemically impossible.  Perhaps Fe doesn't stick well to the
very small carbonaceous grains, instead depleting onto  
a silicate population extending down only to 
$\approx 10$ to $15 \Angstrom$.  Alternatively, there might be another 
population of very small iron-rich grains.
There are well-known oxides which cannot be ruled
out to date, but which have spectral features in the 15 to $30 \micron$
range.

\acknowledgements
This research was supported in part by NSF grants
AST-9219283 and AST-9619429.
We are grateful to J. S. Mathis for suggesting that we include the injection
of material into the ISM from evolved stars and for other helpful comments,
and to R. H. Lupton for the availability of the SM plotting package.

\begin{figure}
\epsscale{1.00}
\plotone{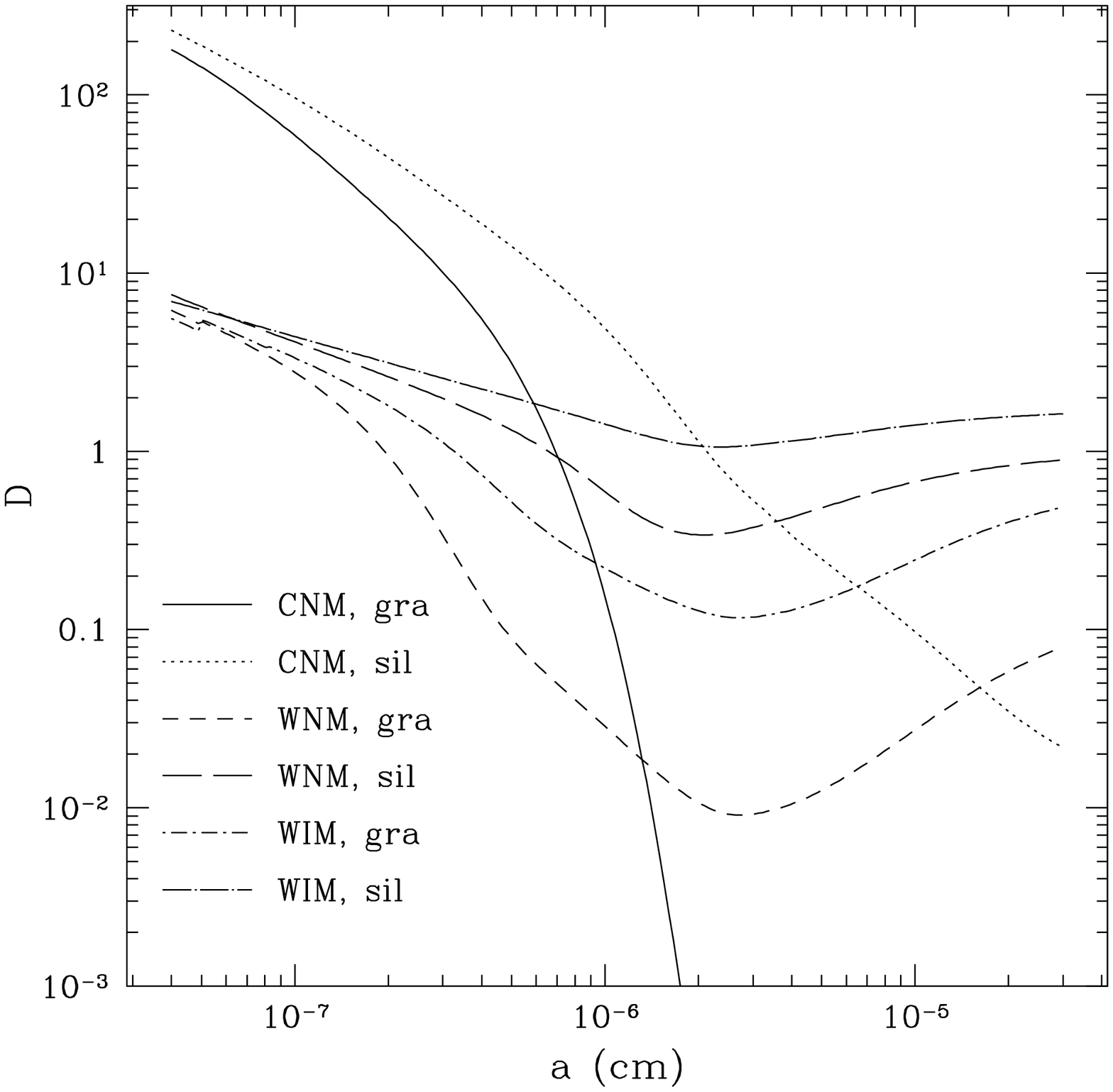}
\caption{
        The enhancement factor, $D(a)$, for the collision of ions with
$Z_i =+1$ with grains.  For graphite grains in the CNM, $D$ drops to
$4 \times 10^{-12}$ at $a = 3 \times 10^{-5} \cm$.
        }
\end{figure}
\begin{figure}
\epsscale{1.00}
\plotone{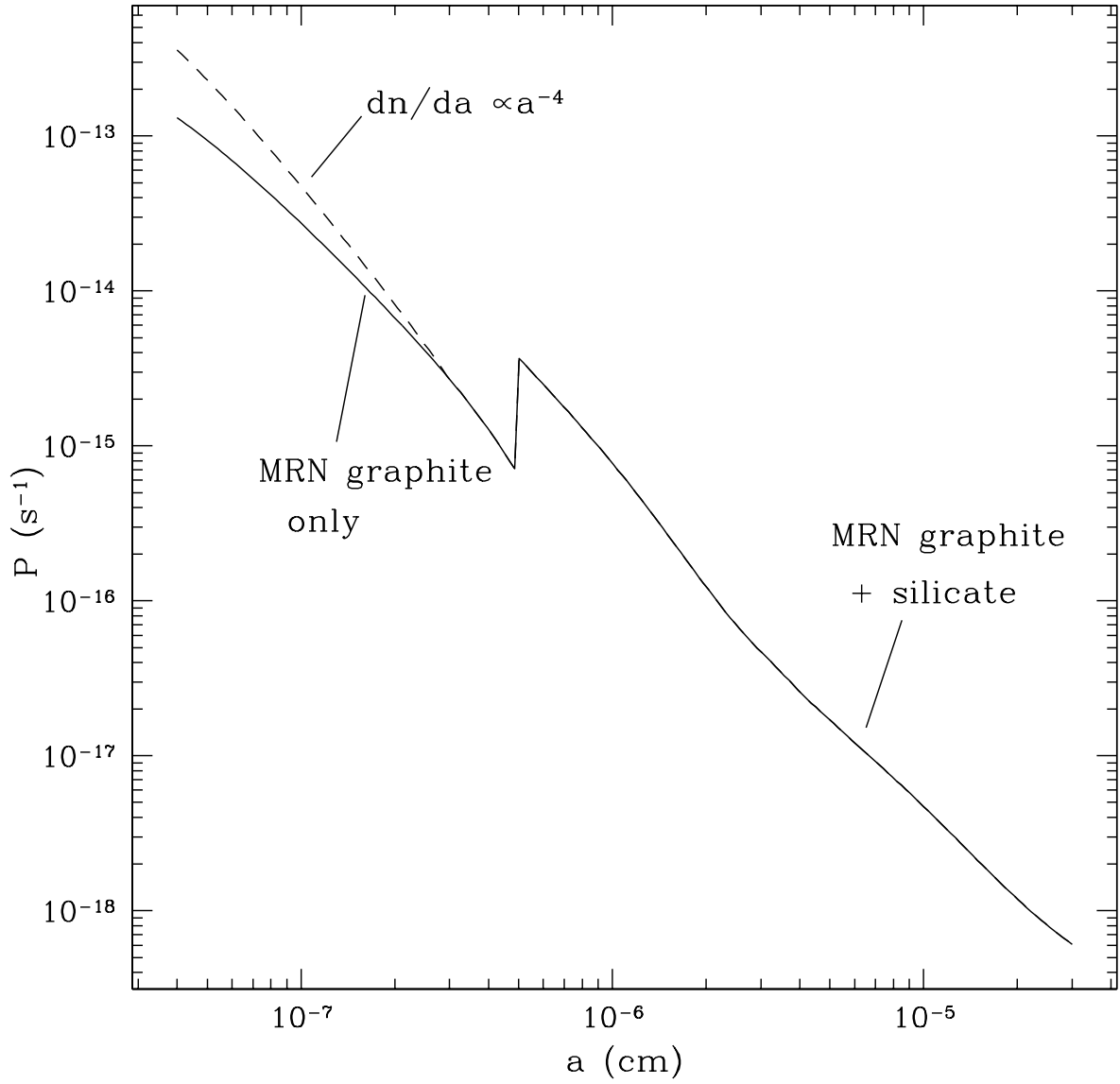}
\caption{
        $P(a)$, showing the contribution to depletion as a function of grain
size, for the CNM.  The discontinuity at $50 \Angstrom$ is due to the 
population
of silicate grains with $a > 50 \Angstrom$.  The lower curve is for grain size
distribution ({\it i}) and the upper curve for distribution ({\it ii}).
        }
\end{figure}
\begin{figure}
\epsscale{1.00}
\plotone{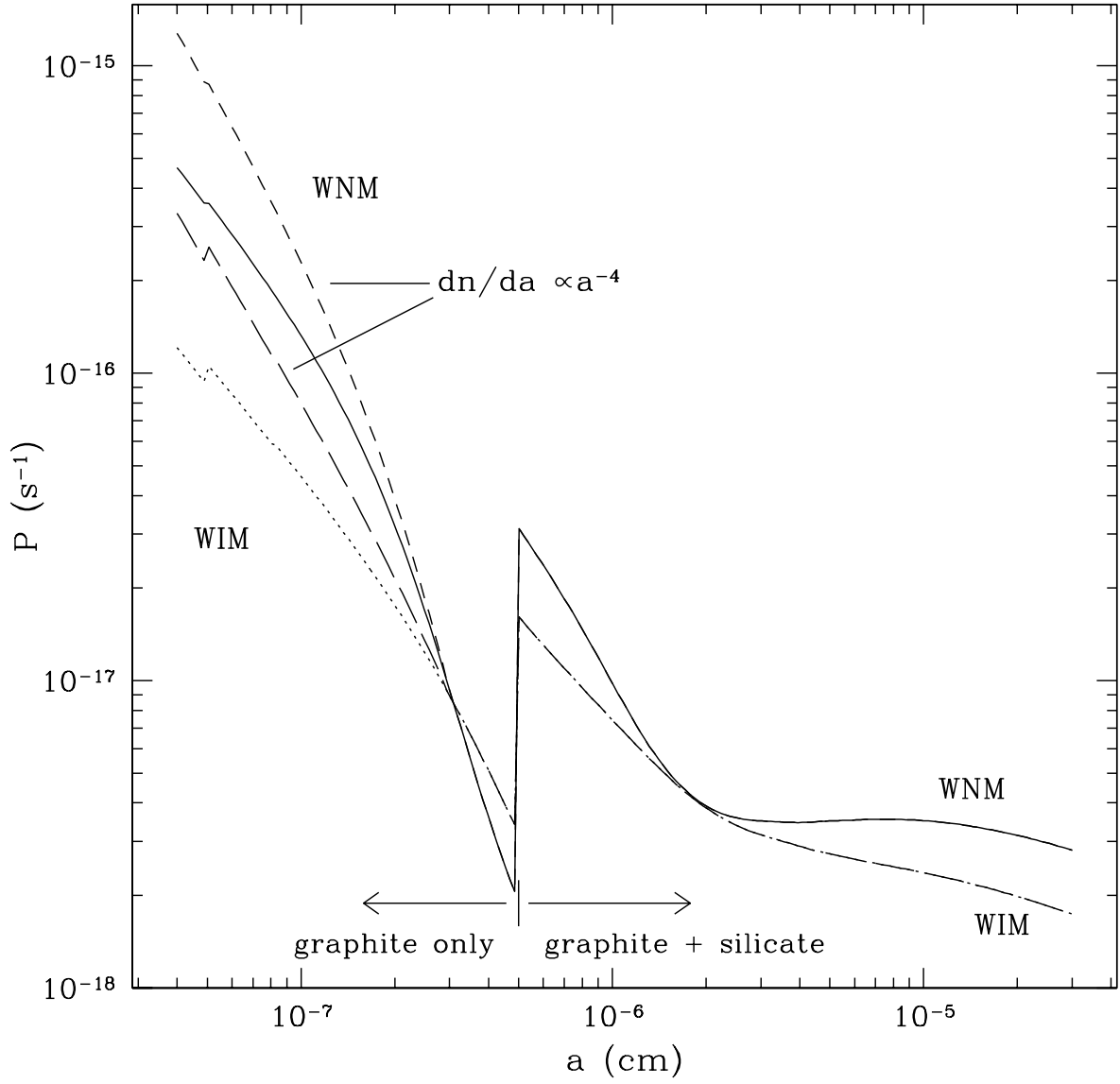}
\caption{
        Same as Figure 2, for the WNM and WIM.
        }
\end{figure}
\begin{figure}
\epsscale{1.00}
\plotone{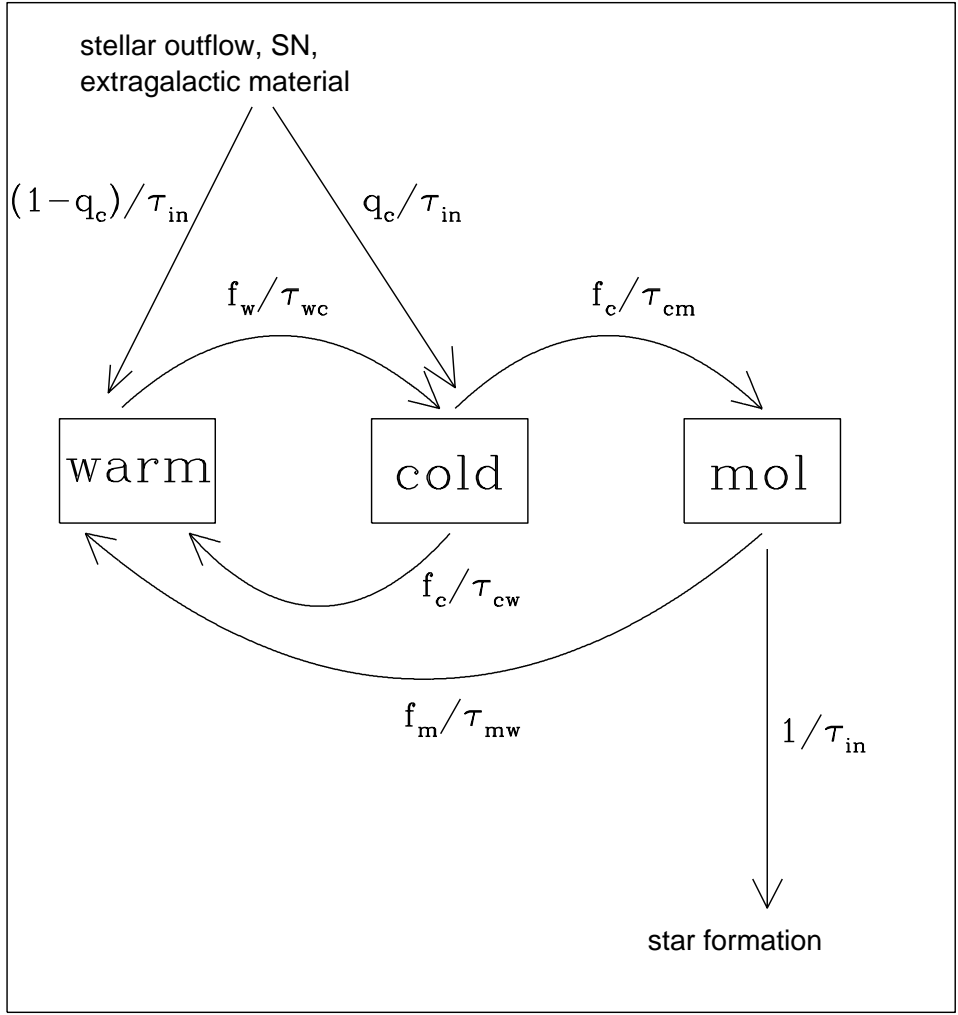}
\caption{
	Mass transfers between warm, cold, and molecular phases.  Each arrow
is labelled by the mass transfer rate divided by the total mass of the ISM.
        }
\end{figure}

\begin{deluxetable}{ccccc}
\tablecaption{Idealized phases for interstellar matter \label{tph1}}
\tablehead{
\colhead{quantity}&
\colhead{CNM}&
\colhead{WNM}&
\colhead{WIM}&
\colhead{MC}
}
\startdata
$\nH (\cm^{-3})$          &30     &0.4    &0.1    &1000   \nl
$T (\K)$                  &100    &6000   &8000   &20     \nl
$x_e$                     &0.0015 &0.1    &0.99   &...      \nl
\enddata
\end{deluxetable}

\begin{deluxetable}{cccc}
\tablecaption{Input Values \label{tph2}}
\tablehead{
\colhead{quantity}&
\colhead{w}&
\colhead{c}&
\colhead{m}
}
\startdata
$\tau_{\rm{a,i}} \, (\rm{yr})$ ({\it i})
&$1.1 \times10^8$  &$4.4 \times 10^5$  &...   \nl
$\tau_{\rm{a,i}} \, (\rm{yr})$ ({\it ii})
&$6.0 \times10^7$  &$2.1 \times 10^5$  &...   \nl
$\tau_{\rm{a,i}} \, (\rm{yr})$ (MRN)  &... &...  &$5.6 \times 10^6$  \nl
$f_{\rm{i}}$                  &0.2   &0.3  &0.5  \nl
$\delta_{\rm{i}}(\rm{Ti})$  &$5 \times 10^{-2}$ &$1 \times 10^{-3}$ &...  \nl
\enddata
\tablecomments{
        For grain size distribution ({\it i}),
the adopted value of $\tau_{\rm{a,w}}$ is intermediate
between the values of $1.0 \times 10^8 \, \rm{yr}$ for the WNM and $2.8 
\times 10^8 \, \rm{yr}$ for the WIM.  For distribution ({\it ii}), the 
corresponding values are $4.9 \times 10^7 \, \rm{yr}$ for the WNM and $1.5
\times 10^8 \, \rm{yr}$ for the WIM.
        }
\end{deluxetable}

\begin{deluxetable}{ccccccccc}
\tablecaption{Exemplary Solutions \label{solns}}
\tablehead{
\colhead{case}&
\colhead{dist}&
\colhead{$\delta_{\rm{in}}$}&
\colhead{$\tau_{\rm{d}}$}&
\colhead{$\tau_{\rm{wc}}$}&
\colhead{$\tau_{\rm{cm}}$}&
\colhead{$\tau_{\rm{cw}}$}& 
\colhead{$\tau_{\rm{mw}}$}& 
\colhead{$\delta_{\rm{m}}$}  
}
\startdata
A& ({\it i}) 
&0.1 &$1.4 \times 10^9$ &$1.5 \times 10^7$ &$4.5 \times 10^7$ &$4.7 \times
10^7$ &$8.7 \times 10^7$ &$5.9 \times 10^{-4}$ \nl
B& ({\it ii})
&0.1 &$6.3 \times 10^8$ &$7.3 \times 10^6$ &$2.4 \times 10^7$ &$2.0 \times
10^7$ &$4.3 \times 10^7$ &$6.1 \times 10^{-4}$ \nl
C& ({\it ii})
&1.0 &$1.6 \times 10^9$ &$7.6 \times 10^6$ &$2.4 \times 10^7$ &$2.1 \times
10^7$ &$4.4 \times 10^7$ &$6.1 \times 10^{-4}$ \nl
\enddata
\tablecomments{
For $\delta_{\rm{w}} = 5 \times 10^{-2}$, $\delta_{c}
=1 \times 10^{-3}$, sticking coefficient $s = 1$,
mass number A $=48$ (Ti), $q_{\rm{c}} =0.1$, $g_{\rm{c}}=0.05$,
$\tau_{\rm{in}} = 1 \times
10^9 \, \rm{yr}$, and $\tau_{\rm{d,m}} = 1 \times 10^{10} \, \rm{yr}$.  
All timescales are in yr.
        }
\end{deluxetable}

\begin{deluxetable}{cccccccc}
\tablecaption{Required Refreshment Timescales \label{compt}}
\tablehead{
\colhead{M}&
\colhead{$\log x_{\rm{cos}}$}&
\colhead{$\delta_{\rm{w}}$}&
\colhead{$\delta_{\rm{c}}$}&
\colhead{$\tau_{\rm{a,c}}^{\rm{req}}$}&
\colhead{$\frac{\tau_{\rm{re}}^{\rm{max}}(10 \Angstrom)}{\beta - \gamma}$}&
\colhead{$\gamma(\rm{M})$}&
\colhead{$\tau_{\rm{mw}}$}
}
\startdata
Mg &-4.42 &0.13 &$2.8 \times 10^{-2}$ &$6.0 \times 10^6$ 
&$3.1 \times 10^8$ &$7.6 \times 10^{-2}$ &$9.0 \times 10^7$ \nl
Si &-4.45 &0.30 &$4.9 \times 10^{-2}$ &$1.3 \times 10^7$ 
&$4.1 \times 10^8$ &$5.6 \times 10^{-2}$ &$3.3 \times 10^8$ \nl
Fe &-4.49 &$5.6 \times 10^{-2}$ &$5.4 \times 10^{-3}$ &$1.0 \times 10^6$ 
&$3.2 \times 10^8$ &$7.4 \times 10^{-2}$ &$4.4 \times 10^7$ \nl
Ca &-5.66 &... &$1.9 \times 10^{-4}$ &$5.5 \times 10^4$ 
&$7.3 \times 10^9$ &$3.3 \times 10^{-3}$ &... \nl
Ni &-5.75 &$3.1 \times 10^{-2}$ &$1.8 \times 10^{-3}$ &$3.4 \times 10^5$ 
&$5.9 \times 10^9$ &$4.1 \times 10^{-3}$ &$2.9 \times 10^7$ \nl
Cr &-6.32 &$8.5 \times 10^{-2}$ &$5.2 \times 10^{-3}$ &$1.0 \times 10^6$ 
&$2.2 \times 10^{10}$ &$1.1 \times 10^{-3}$ &$6.6 \times 10^7$ \nl
Ti &-7.07 &$5.0 \times 10^{-2}$ &$1.0 \times 10^{-3}$ &$2.0 \times 10^5$ 
&$1.3 \times 10^{11}$ &$1.8 \times 10^{-4}$ &$3.7 \times 10^7$ \nl
\enddata
\tablecomments{
The second, third, and fourth 
columns are taken from Savage and Sembach (1996), Table 5.  All timescales 
are in yr.  The depletions $\delta_{\rm{c}}$ and $\delta_{\rm{w}}$ are for
the line of sight to $\zeta$ Oph.  $\tau_{\rm{a,c}}^{\rm{req}}$ is the 
accretion timescale required to yield the observed depletions for a 
destruction timescale $\tau_{\rm{d}} = 6 \times 10^8 \, \rm{yr}$. 
The depletion $\delta_
{\rm{w}}$ for Ca in the warm cloud is unknown.  We assumed that $(1- \delta_
{\rm{w}})/(1-\delta_{\rm{c}})$ is the same for Ca as for Ti in computing
$\tau_{\rm{a,c}}^{\rm{req}}$.
$\tau_{\rm re}^{\rm max}(10\Angstrom)$ 
is the maximum allowed refreshment timescale for $a=10\Angstrom$ carbonaceous
grains,
where $\beta$ is the maximum allowed number of metal atoms per carbon atom,
and $\gamma$ is the number of metal atoms per carbon in ``fresh''
carbonaceous grains.
        }
\end{deluxetable}

\begin{deluxetable}{ccc}
\tablecaption{IR bands for Fe oxides \label{irobs}}
\tablehead{
\colhead{mineral}&
\colhead{IR band position}&
\colhead{reference}
\\
\colhead{}&
\colhead{$\micron$}&
\colhead{}
}
\startdata
hematite ($\alpha$-Fe$_2$O$_3$) &9, 18, 21, 30 &Koike et al. 1981 \nl
magnetite (Fe$_3$O$_4$) &17, 25 &Koike et al. 1981 \nl
w\"{u}stite (FeO) &19.9 &Henning et al. 1995 \nl
\enddata
\tablecomments{
We suspect that the $9 \micron$ feature in the hematite spectrum is due to 
an impurity.
}
\end{deluxetable}

\end{document}